\documentclass[pra, twocolumn, final, reprint, amsmath, amssymb, superscriptaddress, showpacs, a4paper, aps, 10pt, hidelinks, numbers, sort&compress, floats, balancelastpage]{revtex4-1}

\usepackage{graphicx, color} 
\usepackage[normalem]{ulem}          

\usepackage[utf8]{inputenc}

\newcommand{\1}{\ensuremath{\left|1 \right\rangle}}

\definecolor{britishracinggreen}{rgb}{0.0, 0.26, 0.15}
\definecolor{bulgarianrose}{rgb}{0.28, 0.02, 0.03}
\definecolor{darkred}{rgb}{0.90,0,0}
\definecolor{darkgreen}{rgb}{0,0.60,.2}
\definecolor{darkblue}{rgb}{0,0,1}
\definecolor{orange}{cmyk}{0,0.6,0.8,0}
\definecolor{lightblue}{rgb}{0.3,0.5,1}
\definecolor{lightgreen}{rgb}{0.4,0.80,.4}

\newcommand{\beq}{\begin{equation}}
\newcommand{\eeq}{\end{equation}}
\newcommand{\bei}{\begin{itemize}}
\newcommand{\eei}{\end{itemize}}
\newcommand{\ben}{\begin{enumerate}}
\newcommand{\een}{\end{enumerate}}

\begin{document}

\title{Critical Josephson current in BCS-BEC crossover superfluids}


\author{M. Zaccanti} 
\affiliation{Istituto Nazionale di Ottica del Consiglio Nazionale delle Ricerche (INO-CNR), 50019 Sesto Fiorentino, Italy}
\affiliation{\mbox{LENS and Dipartimento di Fisica e Astronomia, Universit\`{a} di Firenze, 50019 Sesto Fiorentino, Italy}}
\author{W. Zwerger}
\affiliation{Physik-Department, Technische Universit$\ddot{a}$t M$\ddot{u}$nchen, D-85748 Garching, Germany}


\begin{abstract}
We develop a microscopic model to describe the Josephson tunneling between two superfluid reservoirs of ultracold fermionic atoms 
which accounts for the dependence of the critical current on both the barrier height and the interaction strength along the crossover from BCS to BEC.
Building on a previous study [F. Meier \& W. Zwerger, Phys. Rev. A, \textbf{64} 033610 (2001)] of weakly-interacting bosons, we derive analytic results 
for the Josephson critical current at zero temperature for homogeneous and trapped systems at arbitrary coupling.
The critical current exhibits a maximum near the unitarity limit which arises from the competition between the increasing condensate fraction
and a decrease of the chemical potential along the evolution from the BCS to the BEC limit.
Our results agree quantitatively with numerical simulations and recent experimental data. 
\end{abstract}

\maketitle

\section{INTRODUCTION}
Josephson weak links \cite{Josephson1962,Anderson1963} represent a paradigmatic tool to investigate phase 
coherence between coupled superfluids  \cite{Anderson1964, Anderson1966}, and they have been realized in a variety of different 
setups: From solid state systems \cite{Likharev1979}, where the Josephson effect has played a crucial role, e.g. to unveil the 
$d$-wave nature of pairing in high temperature superconductors \cite{VanHarlingen1995, Tsuei2000}, to neutral superfluids of 
liquid Helium \cite{Davis2002,Varoquaux2015} and weakly interacting Bose condensates of ultracold atoms 
\cite{Smerzi1997,Zapata1998,Meier2001,Cataliotti2001,Albiez2005,Schumm2005,Levy2007,LeBlanc2011,Jendrzejewski2014,Eckel2014,
Trenkwalder2016,Spagnolli2017}.
More recently, both theoretical \cite{Spuntarelli2007, Salasnich2008a,Ancilotto2009,Adhikari2009,Spuntarelli2010,Zou2014} 
and experimental \cite{Stadler2012,Husmann2015,Valtolina2015,Burchianti2018}  efforts have been targeted to investigate 
the Josephson dynamics in the context of ultracold superfluids of strongly interacting fermion pairs. 
The so-called BCS-BEC crossover \cite{Zwerger2012,Zwerger2016}, which relies on exploiting 
Feshbach resonances between two different hyperfine states of ultracold fermions \cite{Chin2010}, 
offers the unique possibility to investigate, within a single physical system, Josephson tunneling from the standard 
BCS limit of weakly bound fermion pairs to a Bose Einstein condensate of tightly bound molecules, 
including the scale invariant unitary Fermi gas at infinite scattering length as an intermediate phase. 

In contrast with their bosonic counterparts, crossover superfluids are a challenging system to explore. On the experimental side, 
the strong interactions require, for coherent Josephson dynamics to be unveiled, the imprinting of thin barrier potentials on the sample, 
with thicknesses on the order of very few interparticle spacings \cite{Valtolina2015,Burchianti2018}. 
On the theoretical side, the crossover regime poses a difficult many-body problem for which many observables are accessible only
by approximate methods \cite{Zwerger2016}. In particular, microscopic calculations of the Josephson currents based on 
the solution of  Bogoliubov de Gennes equations (BdGE) have been reported in Refs. \cite{Spuntarelli2007,Spuntarelli2010,Zou2014}. 
Although this approach provides a qualitatively correct description of the entire crossover region, it is quantitatively reliable only in the BCS limit.
An alternative strategy relies on the zero-temperature extended Thomas-Fermi models (ETFM)
\cite{Salasnich2008a,Ancilotto2009,Adhikari2009,Zou2014}. These methods are based on a generalized Gross-Pitaevskii equation for the 
complex order parameter $\psi(\textbf{r})$ of bosonic pairs with mass $M=2m$ which tunnel across a barrier of effective height 
$V_0=2 V_{0f}$. Here $m$ and $V_{0f}$ denote the mass and barrier height of a single fermion.
 The pair density is linked to the total fermion density $n=n_{\uparrow}+n_{\downarrow}$ by $|\psi(\textbf{r})|^2 = n(\textbf{r})/2$ 
 (for more details, see e.g. Ref. \cite{Zou2014}).
 A key ingredient of the EFTM non-linear equation is the local chemical potential of pairs $\mu_B$, which is related to the chemical potential 
 $\mu(n, a)$ of a uniform Fermi gas with total density $n$ and interspecies scattering length $a$ by 
 $\mu_B(n, a) = 2 \mu(n, a) + |\epsilon_b|\equiv 2 \mu_F(n, a)$. Here, $\epsilon_b$ is  the binding energy of one dimer 
 in the vacuum ($\epsilon_b=0$ for $a<0$). 
 The advantage of this framework, compared with the mean field BdGE approach, is that  $\mu(n, a)$ is an independent, purely 
 thermodynamic input parameter for the calculation of the Josephson current. Hence, the correct chemical potential throughout 
 the crossover can be accounted for by including the corresponding results obtained from independent Quantum Monte-Carlo calculations, 
 see e.g. Ref. \cite{Astrakharchik2004}.
The EFTM method then properly describes the Josephson effect in the BEC limit. It fails
even qualitatively, however, when extended to the unitary and BCS regimes \cite{Zou2014,Valtolina2015}.
In particular, for fixed barrier heights, ETFM calculations yield a monotonically 
increasing Josephson current when moving from the BEC to the BCS limit. By contrast, the critical current is found to reach a maximum 
value near unitarity both experimentally \cite{Valtolina2015,Burchianti2018} and also in the numerical BdGE calculations  
\cite{Spuntarelli2007,Spuntarelli2010,Zou2014}.  Therefore, developing a unified model which provides a quantitatively correct description 
of the Josephson effect throughout the BCS-BEC crossover represents a relevant and timely problem, which is addressed in the present study. 

Our work builds upon a previous microscopic description of the Josephson dynamics between two weakly interacting Bose Einstein condensates by one of us \cite{Meier2001}. In this framework, the Josephson coupling between two superfluids connected by a high potential barrier is evaluated perturbatively in terms of a tunneling Hamiltonian for bosonic particles. To leading order, this approach gives rise to a critical current which is linearly proportional to the tunneling amplitude of condensed bosons, with a prefactor that characterizes the thermodynamics of the superfluid reservoirs. For lower barriers, the phase-dependent term in their coupling energy includes also a second order correction that involves the excitation of Bogoliubov phonon modes above the condensate, and which increases the maximum Josephson current across the barrier. 
Our aim in the following is to show that a straightforward extension of this model to neutral fermionic superfluids, which exhibit gapless Bogoliubov modes all along the BCS-BEC crossover, quantitatively reproduces both the results of numerical simulations and also the experimental data for the critical current over a wide range of parameters. 
Surprisingly, this holds even beyond the weak tunneling limit considered in the underlying microscopic model. We emphasize that by construction our framework does not account for the effects of real pair-breaking processes, which are irrelevant in the deep tunneling limit since the emerging Josephson coupling energy is then always much smaller than the gap for creating a single fermion excitation. Moreover, we stress that the present work does not intend to provide a description of the full current-voltage characteristic for a finite chemical potential difference $\Delta \mu \neq 0$ between the reservoirs.  Instead, our sole aim here is to determine the critical current which sets the width of the "zero-voltage" branch for arbitrary coupling strength and zero temperature.

Our results suggest that, throughout the BCS-BEC crossover, the critical Josephson current across a tunnel junction 
can be evaluated analytically solely on the basis of (i) the single-particle transmission amplitude for a pair at the relevant 
chemical potential, and (ii) bulk properties of the superfluid at equilibrium, specifically, the condensate density 
and the pair chemical potential. Most importantly, our approach provides a straightforward 
explanation for the experimentally observed maximum of the critical current density near unitarity 
which arises from two competing effects: the increase of the condensate fraction and the concurrent decrease of the 
chemical potential as a function of the dimensionless coupling strength $1/k_Fa$. 
The identification of the unitary gas as the most stable fermionic superfluid \cite{Combescot2006,Miller2007,Spuntarelli2007,Watanabe2009,Weimer2015} 
thus appears from a viewpoint which applies uniformly over the whole regime from weak to strong coupling, with no need to invoke different 
dissipation mechanisms, such as pair breaking on the BCS and phonon-like excitations on the BEC side \cite{Combescot2006,Spuntarelli2007,Zou2014}.

\section{JOSEPHSON CURRENT FROM BCS TO BEC} 
\subsection{First order Josephson current: BEC versus BCS results} 

\begin{figure}[t!]
\begin{center}
\includegraphics[width= 6.5cm]{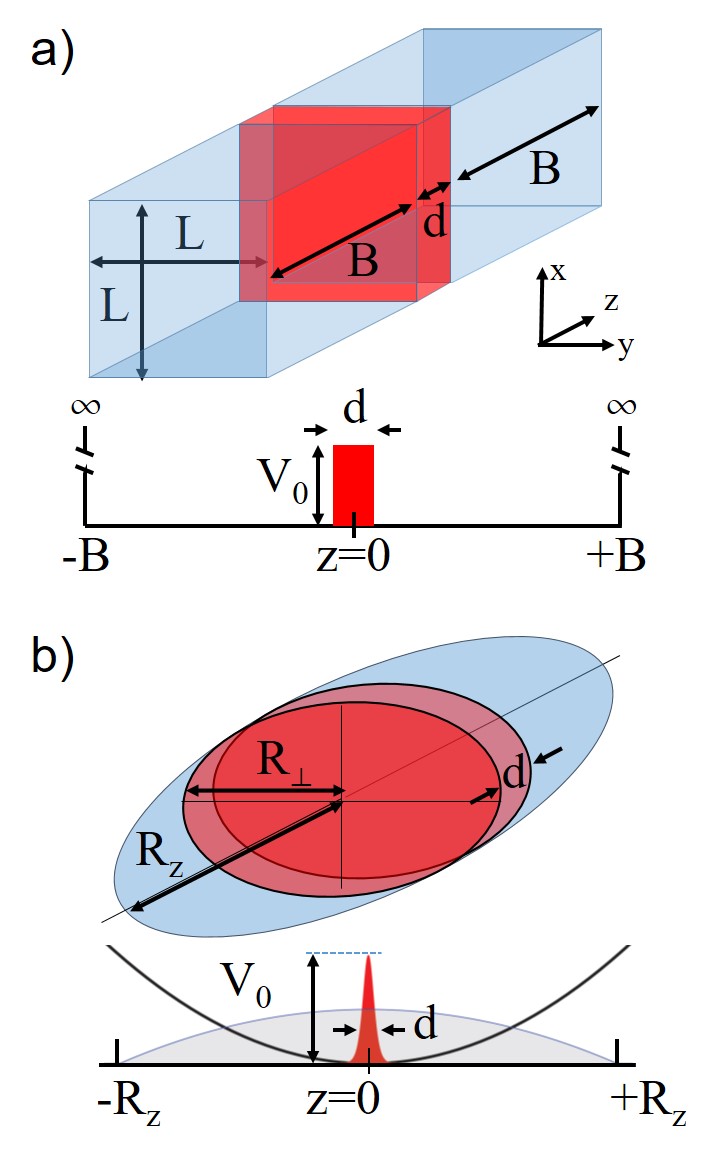}
\caption{Sketch of the three-dimensional Josephson junction setups considered in this work. a) Model geometry relevant for Refs. \cite{Meier2001,Zou2014}, where a  rectangular barrier separates two symmetric homogeneous reservoirs. b) Geometry of the Josephson junction experimentally investigated in Refs. \cite{Valtolina2015,Burchianti2018} where a Gaussian barrier bisects a superfluid confined within a harmonic potential. The radial $R_{\perp}$ and axial $R_z$ radii set the effective section and longitudinal size of the two reservoirs, respectively.}
\label{Fig1}
\end{center}
\hskip -50pt
\end{figure}

In the following, we will argue that the results obtained in Ref. \cite{Meier2001} for the Josephson current between 
two weakly interacting BECs can be extended in a straightforward manner to fermionic superfluids. This will 
provide an analytical result which holds throughout the full crossover from BCS to BEC. 
We start by considering a generic setup of a Josephson junction with the geometry depicted in Fig.\,\ref{Fig1}a, 
where a rectangular barrier connects two homogeneous reservoirs. The results obtained for this idealized case will 
then be generalized to the harmonically trapped situation and Gaussian barriers sketched in Fig.\,\ref{Fig1}b, relevant to 
compare our model predictions with experimental data \cite{Valtolina2015,Burchianti2018}.

For barrier heights $V_0$ which considerably exceed the boson chemical potential $\mu_B$, 
it has been shown in Ref. \cite{Meier2001} that the Josephson currents can be evaluated analytically 
in a systematic expansion in powers of an effective transfer Hamiltonian $\hat{H}_T$. 
In leading order, this gives rise to a non-dissipative Josephson current between the two superfluid reservoirs 
with a sinusoidal current-phase relation $I(\varphi)=I_c \sin{\varphi}$, where  $\varphi$ is the difference between the 
condensate phases across the junction \cite{footnote}. Assuming a homogeneous situation with transverse area $A=L^2$, 
the associated critical current density 
\begin{eqnarray}
\hbar j_{c} =\frac{\hbar I_c}{A} = 2B \, t_{cc}(\mu_B) \, n_c 
\label{jc1}
\end{eqnarray}
can be expressed as the product of the transfer matrix element $t_{cc}(\mu_B)$, 
associated with coherent tunneling of bosons at an energy equal to their equilibrium chemical potential $\mu_B$, 
times the associated condensate density $n_c$. Formally, Eq. (\ref{jc1}) also contains the total longitudinal size 
$2B$ of the system which however drops out, as expected. 
Indeed, for the junction geometry of Fig.\,\ref{Fig1}a, the matrix element $t_{cc}$ has been explicitly evaluated in Ref. \cite{Meier2001}, 
and we conveniently recast it here as 
\begin{eqnarray}
t_{cc}(\mu_B)= \frac{\vert t \vert (\mu_B)}{4k(\mu_B)B}\, \mu_B\,.
\label{eq2}
\end{eqnarray}
Here $k(\mu_B)=\sqrt{2M\mu_B}/\hbar$ is the wave vector of a boson with mass $M$ and chemical potential $\mu_B$, 
while $\vert t \vert (\mu_B)$ is the associated single-particle transmission amplitude.
The critical current density can be therefore expressed in the simple form
\begin{eqnarray}
\hbar j_{c}  =  \frac{\mu_B n_c}{2k(\mu_B)}\, \vert t \vert (\mu_B)\,, 
\label{jc2}
\end{eqnarray}
which shows that inter-particle interactions affect $j_{c}$ only through bulk properties of the superfluid.
Specifically, they involve  the chemical potential $\mu_B$ and the condensate density $n_c$ which, 
to lowest order in the interactions between the pairs, coincides with the full Bose density. 
It is important to notice that the microscopic tunneling amplitude $ \vert t \vert (\mu_B)$ is the one for a \textit{single} boson.
It depends on the interaction strength only to the extent that this determines the relevant energy at which the tunneling process occurs.
As discussed in Ref. \cite{Meier2001}, the separation between single particle and many-body properties within Eq. (\ref{jc2}) 
originates from the underlying assumption that the barrier height $V_0$ greatly exceeds the chemical potential $\mu_B$. 
Under these conditions, our major claim in the following is that Eq. (\ref{jc2}) correctly describes the Josephson current at \textit{any} coupling strength.

The fact that Eq. (\ref{jc2}) may hold even for fermionic superfluids throughout the crossover from the BCS to the BEC limit 
is supported by the quite remarkable observation that, upon a proper identification of the variables, 
this result essentially coincides with the well known expression for the critical current between two BCS superconductors. 
Within BCS theory, the critical current of a Josephson junction in the tunneling limit is known to be equal to the current flowing in 
the normal state at a finite applied voltage $eV=\pi\Delta_0/2$, where $\Delta_0$ is the energy gap at 
zero temperature \cite{Josephson1962,Ambegaokar1963,Anderson1964}.
By evaluating the normal state conductance per unit area, which scales linearly with the number of discrete transverse modes, 
the Ambegaokar-Baratoff (AB) formula for the \textit{pair} current density between two BCS superfluids coupled by a thin tunneling barrier
can be expressed in the form \cite{Spuntarelli2010}
\begin{eqnarray}
\hbar j_c^{AB} = \frac{\pi}{2} \Delta_0 \, \frac{k_F^2}{16\pi^2}\,\vert t \vert^2(\mu_F)
\label{IAB}
\end{eqnarray}   
where $\vert t \vert^2(\mu_F)$ is the transmission probability of a single fermion at the Fermi energy 
$\mu_F\to \epsilon_F$. While at a first glance the expressions in Eqs. (\ref{jc2}) and (\ref{IAB}) appear totally 
disconnected, it turns out that they are essentially equivalent. 
Indeed, in the BCS limit, the condensate density $n_c=\lambda_0\cdot n/2$ is linked to the gap by the relation 
(see e.g. Refs. \cite{Ortiz2005,Salasnich2005})

\begin{eqnarray}
\lambda_0^{\rm BCS} = \frac{3\pi}{8}\,\frac{\Delta_0}{\epsilon_F}
\label{condvsgap}
\end{eqnarray}
where $n=k_F^3/(3\pi^2)$ is the total fermion density and $\lambda_0$ denotes the condensate fraction.
 Hence, Eq. (\ref{IAB}) can be recast in the form
\begin{eqnarray}
\hbar j_c^{AB} = \frac{\mu_F n_c }{2 k_F}\, \vert t \vert^2(\mu_F)
\label{IAB2}
\end{eqnarray}
where the similarity with Eq. (\ref{jc2}) is now apparent. Indeed, noting that $2k_F \sim 2\sqrt{2m\mu_F}/\hbar\to k(\mu_B)$
in terms of the chemical potential $\mu_B=2\mu_F$ of pairs, the two expressions are perfectly equivalent if 
the transmission probability $\vert t \vert^2(\mu_F)$ of a single fermion is replaced by the transmission amplitude
$\vert t \vert (\mu_B)$ of a pair. 

More specifically, it is instructive to consider the thermodynamic prefactor, common to the BCS result Eq. (\ref{IAB2}) and 
the BEC one Eq. (\ref{jc2}) for the critical current density, for a fermionic superfluid at arbitrary coupling. 
This quantity, which encodes the many-body properties of the system, can be suitably expressed in terms of the condensate fraction 
$\lambda_0=2n_c/n$ and the dimensionless chemical potential $\tilde{\mu}=\mu_F/\epsilon_F$ of one fermion, 
which enters into the BEC result Eq. (\ref{jc2}) through the standard mapping  $\mu_B(n, a) = 2 \mu(n, a) + |\epsilon_b|\equiv 2 \mu_F(n, a)$
between the Bose and Fermi chemical potential discussed in the previous section. 
With $v_F=\sqrt{2\epsilon_F/m}$ as the Fermi velocity of the non-interacting Fermi gas one obtains 
\begin{eqnarray}
\frac{j_c}{\vert t \vert (\mu_B)}= \frac{j_c^{AB}}{\vert t \vert^2 (\mu_F)}= \frac{n v_F}{8} \lambda_0 \sqrt{\tilde{\mu}}\,.
\label{general}
\end{eqnarray}
Eq. (\ref{general}) shows that the critical current density, in units of the natural scale set by $(n\, v_F)$, is determined by the product 
of the condensate fraction $\lambda_0$ with the square root of the (normalized) fermion chemical potential $\sqrt{\tilde{\mu}}$.
\begin{figure}[t!]
\begin{center}
\includegraphics[width= 8.5cm]{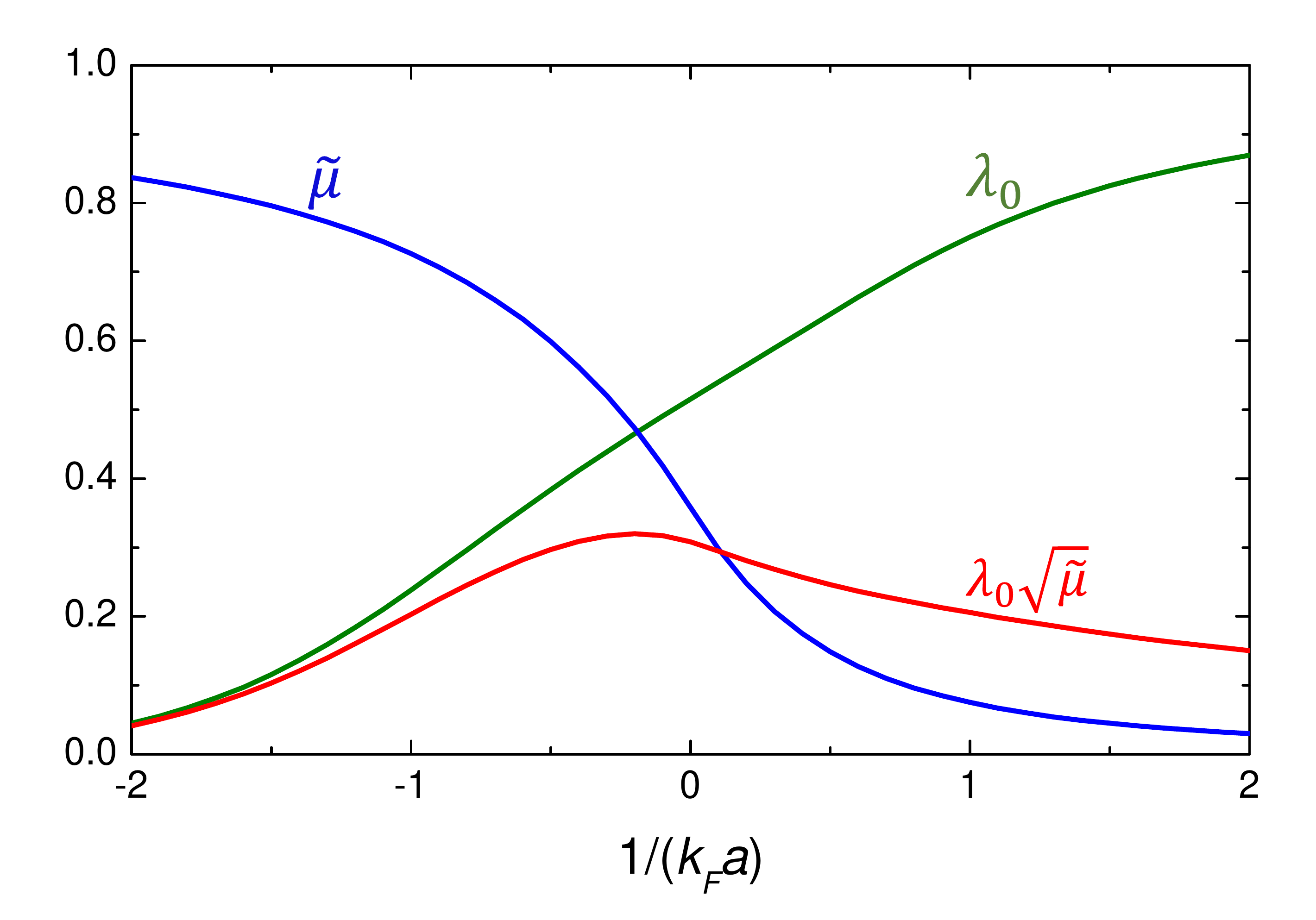}
\vskip -10pt
\caption{Condensate fraction $\lambda_0$ (green line) and normalized chemical potential $\tilde{\mu}=\mu_F/\epsilon_F$ (blue line) 
for three-dimensional crossover superfluids obtained 
within a Luttinger-Ward approach at zero temperature \cite{Haussmann2007}, as a function of the dimensionless coupling strength $1/k_Fa$. 
The product $\lambda_0\, \sqrt{\tilde{\mu}}$ (red line), which determines the dependence of the critical Josephson current on interactions 
according to Eq. (\ref{general}), exhibits a maximum near unitarity.}
\label{mu-lambda}
\end{center}
\hskip -50pt
\end{figure}
The dependence of these two thermodynamic quantities on the coupling strength is shown in Fig. \ref{mu-lambda}, 
which presents the results for $\lambda_0$ (green line) and $\tilde{\mu}$ (blue line) obtained within a Luttinger-Ward approach 
to the BCS-BEC crossover problem developed by~\textcite{Haussmann2007}. Apparently, these two bulk properties exhibit 
an opposite behavior as a function of the coupling strength: while $\tilde{\mu}$  
monotonically decreases from one towards zero upon moving from the BCS to the BEC limit, the condensate fraction $\lambda_0$ 
monotonically increases from exponentially small values in the BCS regime towards unity for a weakly interacting Bose superfluid.
A qualitatively similar behavior is also obtained within a mean field BdGE approach, see e.g. Ref. \cite{RMP2008b} and references therein. 
However, the results of the two theories differ substantially on a quantitative level. In particular, for the gas at unitarity, mean field calculations 
yield a condensate fraction $\lambda_0 \simeq 0.7$ instead of $0.51$, and a chemical potential $\tilde{\mu} \simeq 0.56$ 
instead of the value $0.36$ obtained within the Luttinger-Ward  approach, which agrees quite well with the experimental value 
$\tilde{\mu}=\xi_s=0.37$ for the associated Bertsch parameter, see Ref. \cite{Ku2012}.
 
As a consequence of the competing trends of $\lambda_0$ and $\tilde{\mu}$, the thermodynamic pre-factor in Eq. (\ref{general})  
reaches a maximum near unitarity (see red line in Fig. \ref{mu-lambda}), in qualitative agreement with both experimental \cite{Valtolina2015} 
and numerical \cite{Spuntarelli2007,Zou2014} findings. 
Note, however, that the origin of the non-monotonic behavior in the present theory is quite different from the interpretation 
given by \textcite{Spuntarelli2007} for the case of  weak barriers. There, following a Landau criterion, the rise and fall of the maximum 
Josephson current was ascribed to the existence of two distinct types of critical velocities associated with two different excitation branches: 
pair-breaking on the BCS and phonons on the BEC side, respectively. By contrast, our result Eq. (\ref{general}) shows that in the tunneling regime
this peculiar non-monotonicity emerges from the competition between $\lambda_0$ and $\tilde{\mu}$ uniformly throughout the crossover region.
This conclusion is further  supported by recent experimental studies \cite{Valtolina2015,Burchianti2018}, which unambiguously identified 
vortex rings and phonons, rather than "fermionic" pair-breaking excitations, as the microscopic mechanisms responsible for the breakdown of the 
dissipationless flow, even on the BCS side of the crossover.

In the following, we focus on the difference between the BEC result in Eq. (\ref{jc2}) and the BCS one of Eq. (\ref{IAB2}) that, 
as highlighted in Eq. (\ref{general}), amounts to the ratio between the tunneling amplitude of one pair and the transmission 
probability of one fermion, evaluated at their relevant energies.
Naively, such a quantity could be expected to exhibit an exponential dependence upon the barrier thickness $d$ and the 
normalized height $V_0/\mu_B$, hence causing a large quantitative mismatch between the BCS and BEC results.
However, employing the mapping of the "fermionic" BdGE approach onto the "bosonic" ETFM one \cite{Zou2014}, 
we have verified that this is not the case: The ratio of these two transmission terms is of order unity for a wide range of barrier geometries,
 featuring only a very weak dependence upon the barrier properties and the chemical potential. 
Indeed, applying Eq. (\ref{jc2}) to the BCS limit of the crossover yields a critical current density that matches the 
BCS result Eq. (\ref{IAB2}) within a few $10\%$ accuracy. For instance, over an extremely wide range $k_F d \in [10^{-3},10^2]$
of dimensionless barrier thicknesses, the two terms match one another within $\pm 50\%$, both for rectangular and Eckart 
barriers of relative heights $V_0/\mu_B \in [1,10^4]$.  Within the experimentally relevant regime of barrier widths 
$1 \lesssim k_F d \lesssim 4$ and moderate heights, $V_0/\mu_B \lesssim 2$ \cite{Valtolina2015,Burchianti2018},
the deviations are in fact smaller than $20\%$. In view of these results, in the following we will thus use the extension of the bosonic 
expression Eq. (\ref{jc2}) to arbitrary couplings to describe the critical current of a Josephson junction along the entire BCS-BEC crossover.

\subsection{Second order contributions and extension to inhomogeneous samples} 

In order to benchmark our analytic theory against numerical and experimental results, it is necessary to also 
take into account higher harmonics that contribute to the critical current. In fact, as already mentioned in the previous section, 
the tunneling limit of exponentially small values of $|t|(\mu_B)$ is hard to achieve in practice and realistic barrier heights $V_0$ 
are close to, or only slightly larger than, the chemical potential \cite{Valtolina2015,Burchianti2018}. 
Therefore, higher order terms in the expansion in powers of the transfer Hamiltonian are non-negligible.
In particular, in second order, these lead to an additional non-dissipative contribution 
$I_1\sin{2\varphi}$ to the coherent flow. It has half of the period $\varphi\to\varphi+2\pi$ of the fundamental Josephson 
current and, within our framework, it arises from condensate to non-condensate tunneling processes, involving the excitation of a Bogoliubov phonon.  
By suitably recasting the results of Ref. \cite{Meier2001}, the magnitude of the (negative) second harmonic current density can be linked to the dominant first order term via
\begin{eqnarray}
\vert j_1 \vert=j_{c}
\, \frac{\, \vert t \vert (\mu_B)}{4}\, , 
\label{jsecond}
\end{eqnarray}
which still exhibits the separation between single-particle and many-body properties discussed above for the leading contribution given by Eq. (\ref{jc2}).
It is important to remark that Ref. \cite{Meier2001}, at second order in the transfer Hamiltonian, also predicts normal and phase-dependent 
dissipative currents.  These terms are absent, however, at a vanishing value $\Delta \mu=0$ of the difference in chemical potential 
between the two superfluid reservoirs.  As a result, these contributions are irrelevant for the determination of the maximum Josephson current, 
which is the aim of our present study.

Moreover, in order to test the predictions of our model against recent experimental data \cite{Burchianti2018} and 
BEC results of EFTM numerical simulations \cite{Xhani2019a} available in the literature, it is necessary to 
generalize the results obtained in Eq. (\ref{jc2}) to the experimentally relevant case \cite{Valtolina2015,Burchianti2018} 
of inhomogeneous samples and Gaussian barriers, see Fig.\,\ref{Fig1}b.
In particular, the result in Eq. (\ref{jc2}) for the (first order) critical current density can be easily extended to account for 
an inhomogeneous condensate density profile $n_c(\textbf{r})$ as
\begin{eqnarray}
\hbar j_{c}(x,y) = \int_{-R_z}^{R_z} dz\, n_c(\textbf{r}) \,\mu_B(\textbf{r}) \,\frac{\vert t \vert (\mu_B(\textbf{r}))}{4 \, k(\mu_B)(\textbf{r}) \,R_z}  
\label{jc1gen}
\end{eqnarray}
Further integration of $j_c$ along the transverse directions $(x,y)$ then yields the total current
\begin{eqnarray}
\hbar I_{c} = \int d^3r \, n_c(\textbf{r}) \, \mu_B(\textbf{r}) \,\frac{\vert t \vert (\mu_B(\textbf{r}))}{4 \,  k(\mu_B)(\textbf{r}) \,R_z} \, ,
\label{Ic1gen}
\end{eqnarray}
which obviously recovers the result obtained in the homogeneous case \cite{Meier2001}, where $I_c$ scales linearly with
the junction area $A$.  In the generic inhomogeneous case, Eq. (\ref{Ic1gen}) may be regarded as a sum of transmitted particles over all 
infinitely small volumes centered at position $(r,z)$, each one containing a number $n_c(r,z)$ of condensed bosons, at energy $\mu_B(r,z)$.

By following a similar procedure and recalling Eq. (\ref{jsecond}), one can straightforwardly obtain an analytic expression also  
for the second order contribution to the Josephson current for inhomogeneous samples. While this latter correction is negligible 
in the deep tunneling limit $V_0/\mu_B \gg 1$, it may cause an up to 15\% increase of the maximum supercurrent $I_{Max}$ 
in the experimentally relevant regime $V_0/\mu \sim 1$. This can be easily verified on the basis of the analytic results obtained by 
\textcite{Goldobin2007}, that theoretically investigated how an arbitrary second harmonic term in the current-phase relation of a 
generic  junction affects  the maximum Josephson current.\\

\section{COMPARISON WITH B\MakeLowercase{d}GE NUMERICAL RESULTS FOR THE CROSSOVER REGION}

\begin{figure}[t!]
\begin{center}
\includegraphics[width= 8.5cm]{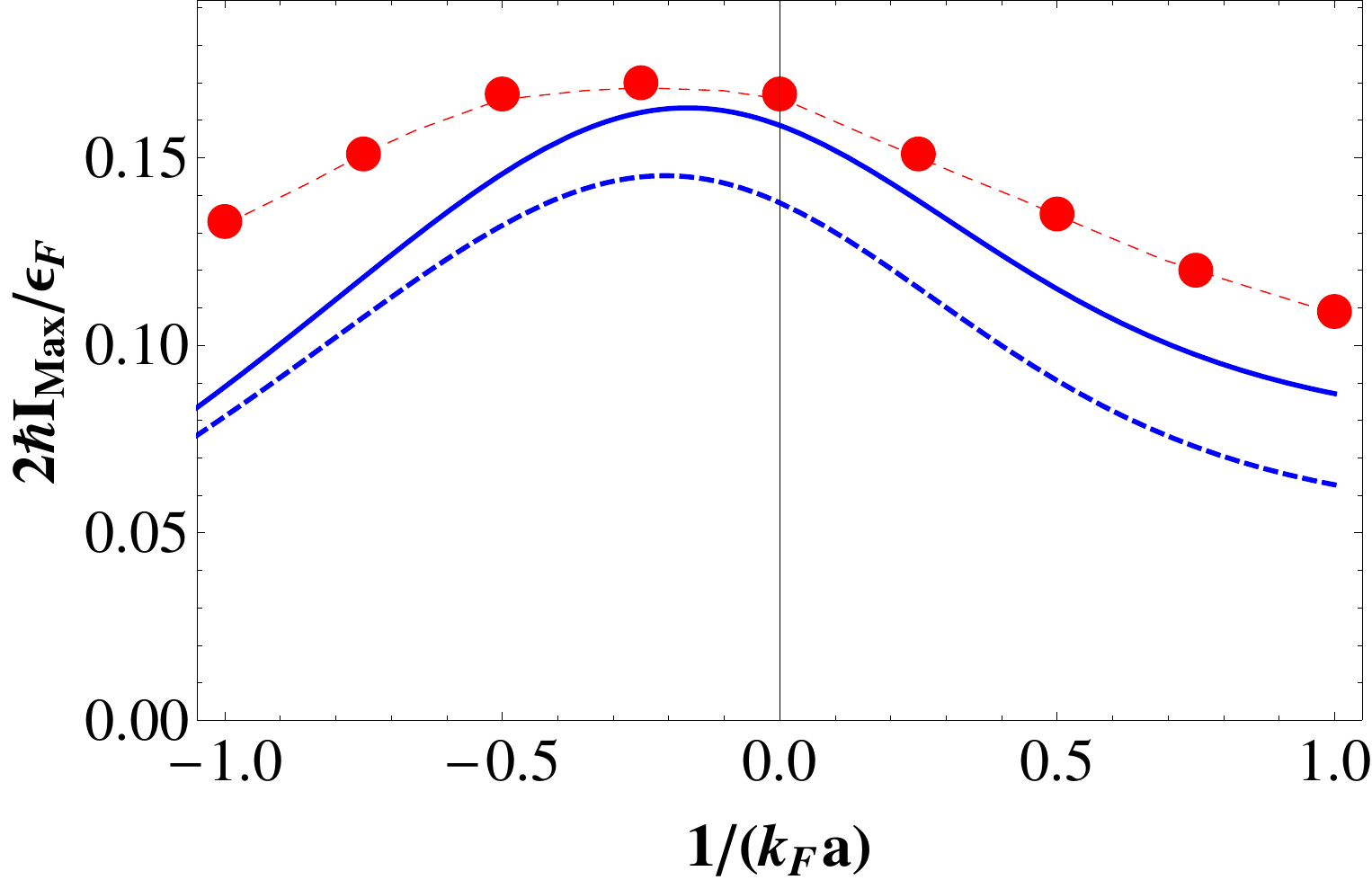}
\caption{
Critical atom current for a Josephson junction in a box throughout the BEC-BCS crossover. 
Red dots are the BdGE results reported in Ref. \cite{Zou2014} for
$V_{0f}/\epsilon_F=5$, $k_F d=0.6$, N=100, $k_F L_{\perp}=13$ and $k_F B=10$. 
The red dashed line is a linear interpolation to the BdGE data.
The dashed blue line is the prediction of our model employing the continuum mean field results \cite{Ortiz2005,RMP2008b} for $\mu_B$ and $n_c$,
whereas the solid blue line accounts for the proper chemical potential of the superfluid in the box geometry \cite{Zou2019}. The difference 
between the two blue lines thus highlights the effect of the confining box on the superfluid chemical potential. }
\label{Fig10Zou}
\end{center}
\hskip -50pt
\end{figure}

In the following, we compare the predictions of our model for a fixed barrier geometry with numerical results based on the 
solution of time-dependent mean field BdGE, reported in Ref. \cite{Zou2014}. 
To this end, we consider  the specific tunnel junction setup of \textcite{Zou2014}, analogous to the one sketched in Fig. \ref{Fig1}a. 
In Fig. \ref{Fig10Zou} the dashed blue line represents the critical fermion current ($2\, I_{Max}$) predicted by our analytic model 
for the barrier and reservoir parameters detailed in the figure legend, and employing the mean field continuum results for chemical potential 
and condensate density, respectively (see Ref. \cite{RMP2008b} and references therein). 

As already emphasized in the previous section,  we obtain a non-monotonic critical current as the interaction is changed from the BCS to the BEC regime.
Compared with the behavior of $\lambda_0\,\sqrt{\tilde{\mu}}$ presented in Fig.\,\ref{mu-lambda}, the shape of $I_{Max}$ 
as a function of $1/k_Fa$ is quantitatively modified owing to the additional energy dependence of $\vert t  \vert(\mu_B)$, 
which monotonically decreases when moving from the BCS towards the BEC limit.
In particular, this causes an overall suppression of the maximum Josephson current, 
more pronounced in the strong attraction limit of tightly bound pairs, where the chemical potential most strongly varies with the coupling strength.  
\begin{figure}[t!]
\begin{center}
\includegraphics[width= 8.5cm]{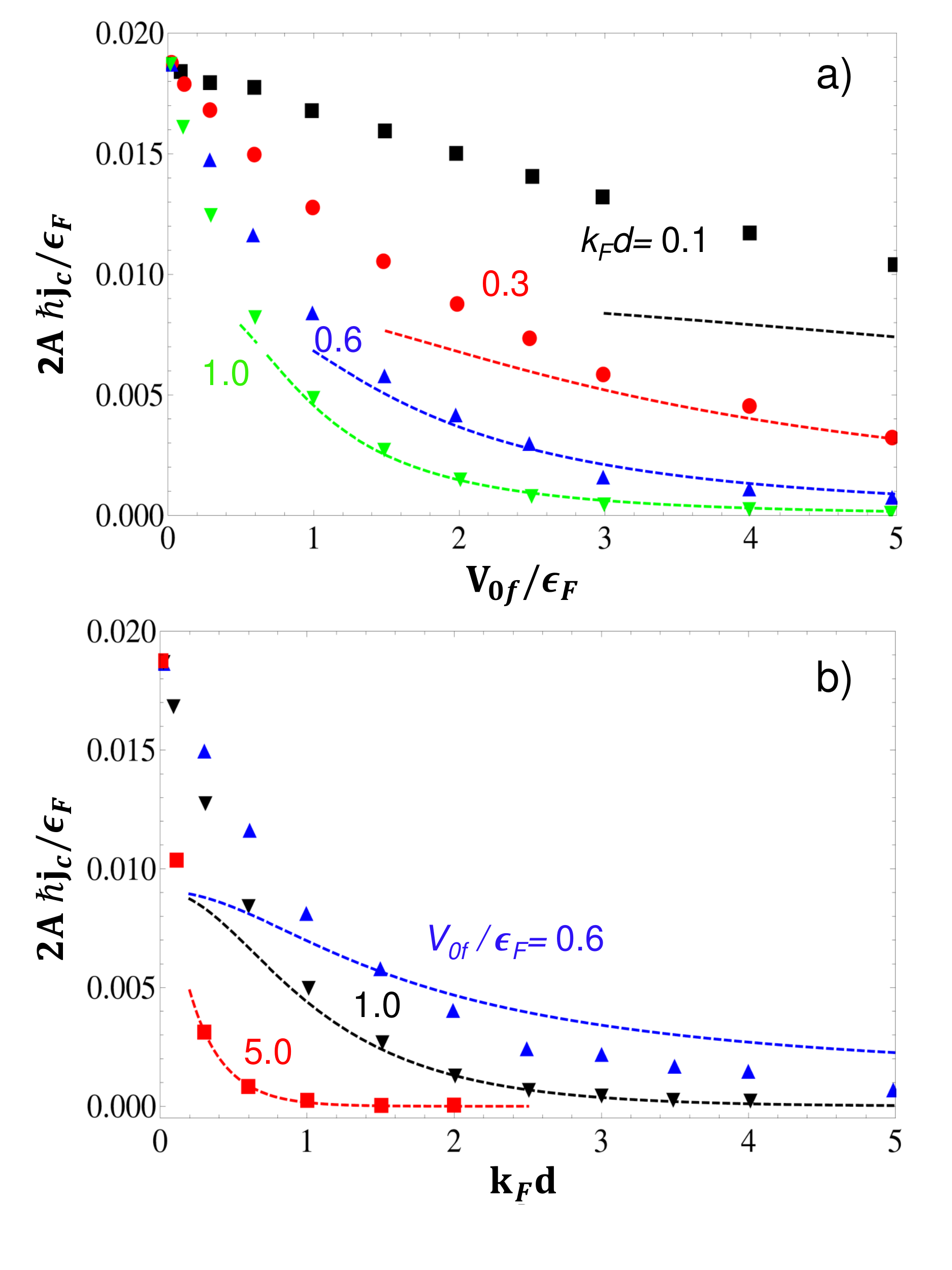}
\vskip -20pt
\caption{Critical current density of a unitary superfluid junction in a box geometry with transverse area $A$. 
a) Symbols are the results reported in Ref. \cite{Zou2014} for the normalized   
fermion current density $2 A \hbar j_c/\epsilon_F$, as a function of barrier height $V_{0f}/\epsilon_F$, obtained for different barrier thicknesses 
as specified in the legend. For all $k_Fd$ values, the dashed lines represent
the predictions of our analytic model from Eq. (\ref{jc2}), accounting for the 
proper chemical potential of the gas \cite{Zou2019} while assuming the condensed fraction obtained by mean field theory in the continuum 
\cite{Ortiz2005,RMP2008b}. The curves also account for second order contributions, see Eq. (\ref{jsecond}) and Ref. \cite{Goldobin2007}.
b) Comparison between BdGE results of Ref. \cite{Zou2019} and our theory model as in panel a) for $2 A \hbar j_c/\epsilon_F$, 
in this case as a function of the barrier thickness and different barrier heights.
}
\label{Fig13Zou}
\vskip -20pt
\end{center}
\end{figure}

Notably, our model appears in reasonable quantitative agreement with the results of BdGE simulations \cite{Zou2014}, see red dots in Fig. \ref{Fig10Zou}. 
In particular, the mismatch between our theory and the numerical results is largely ascribable to finite size effects connected with the small box 
employed in the simulation, strongly affecting the bulk superfluid properties, relative to the continuum ones. 
For instance, the agreement significantly improves when we employ the BdGE chemical potential values obtained within the actual tightly 
confining box \cite{Zou2019}, systematically higher than the continuum result, see blue solid line in Fig. \ref{Fig10Zou}. 
In this case, our theory matches the numerical data with deviations not exceeding 30$\%$ throughout the crossover region.
Additionally, the residual mismatch can be largely attributed to effects not taken into account when evaluating our model predictions: 
First, the small box size may also cause an increased condensate fraction, relative to the continuum case, especially on the BCS side of the crossover. 
Second, the determination of the critical current by solving the time dependent BdGE was obtained by imparting a sizable initial excitation to the system \cite{Zou2014}, yielding an additional kinetic energy contribution to the overall particle energy, neglected in our model. 


The ability of our analytic theory to yield quantitatively accurate Josephson currents in the crossover region is further demonstrated by the 
comparison presented in Fig. \ref{Fig13Zou}. There, we contrast our prediction based on Eq. (\ref{jc2}) for the fermion current density ($2 \hbar j_c$)  
of unitary superfluids with the numerical results obtained by \textcite{Zou2014} solving stationary BdGE, both as a function of the barrier height 
for a few values of the dimensionless barrier thickness (Fig. \ref{Fig13Zou}a), and vice-versa for different barrier heights as a function of the 
thickness (Fig. \ref{Fig13Zou}b). Following the convention of Ref. \cite{Zou2014}, the barrier height is given in the legends for one fermion, $V_{0f}=V_0/2$.
Apparently, our analytic predictions quantitatively reproduce the numerical results remarkably well  
over a wide range of barrier geometries. Indeed, as long as the barrier height exceeds the superfluid chemical potential, sizable deviations are observed 
only for the case of anomalously thin barriers (see black squares in Fig. \ref{Fig13Zou}a), for which our theoretical model 
based on Eq. (\ref{jc2}) would converge to the numerical data only for exceedingly high $V_0$ values. On the other hand and not surprisingly, 
our model fails even at the qualitative level for $V_0/\mu <1$, i.e. out of the tunneling regime, see e.g. blue line and triangles in 
Fig. \ref{Fig13Zou}b (BdGE calculations in the present box geometry yield at unitarity $\mu_F/\epsilon_F \simeq 0.7$ \cite{Zou2019}).

\section{COMPARISON WITH EXPERIMENTS}

Finally, we compare our model predictions with experimental data that were obtained in the Lithium lab at LENS \cite{Burchianti2018}, 
and also with ETFM numerical simulations carried out in the BEC limit \cite{Xhani2019a}. 
To this end, we evaluate the maximum Josephson current $I_{Max}$ accounting for both first and second order contributions \cite{Goldobin2007},
in the case of harmonically trapped samples and Gaussian barriers (see Fig. \ref{Fig1}b), exploiting Eqs. (\ref{Ic1gen}) and (\ref{jsecond}), respectively.
For simplicity, the Gaussian barrier is approximated with an Eckart potential of the form 
\begin{eqnarray}
V(z)=\frac{V_0}{\cosh^{2}(z/d)}
\label{Eckart}
\end{eqnarray}
for which the single-particle transmission probability can be evaluated analytically, see e.g. Ref. \cite{LLbook}.
In particular, it can be verified that a Gaussian profile with $e^{-2}$ waist $w$ can be excellently approximated by an Eckart function with $d\sim 0.6\, w$. 
Inserting the transmission amplitude of the Eckart barrier both in Eq. (\ref{Ic1gen}) for the first order current and in the analogous 
expression for the second harmonic based on Eq. (\ref{jsecond}), and accounting for the relevant superfluid atom number and trap frequencies 
employed by \textcite{Burchianti2018}, we obtain our zero-temperature model predictions for the resulting maximum current $I_{Max}$ \cite{Goldobin2007}. 
For this comparison, we exploited the chemical potential and condensate fraction obtained within the Luttinger-Ward approach 
\cite{Haussmann2007}. An important point to emphasize in this context is that the approach provides a quantitatively precise description of
the thermodynamic properties not only near unitarity, as shown in Fig. \ref{mu-lambda}, but also in the BCS and the deep BEC limit. 
Indeed in the regime $1/k_Fa>2$, the normalized chemical potential $\tilde{\mu}\to k_Fa_{dd}/(3\pi)$ approaches the result associated 
with a dilute gas of Bosons with a dimer-dimer scattering length $a_{dd}=0.59\, a$ which is very close to the exact value $a_{dd}=0.6\, a$ \cite{Petrov2004}.
 
\begin{figure}[t]
\begin{center}
\vskip-10pt
\includegraphics[width= 8.5cm]{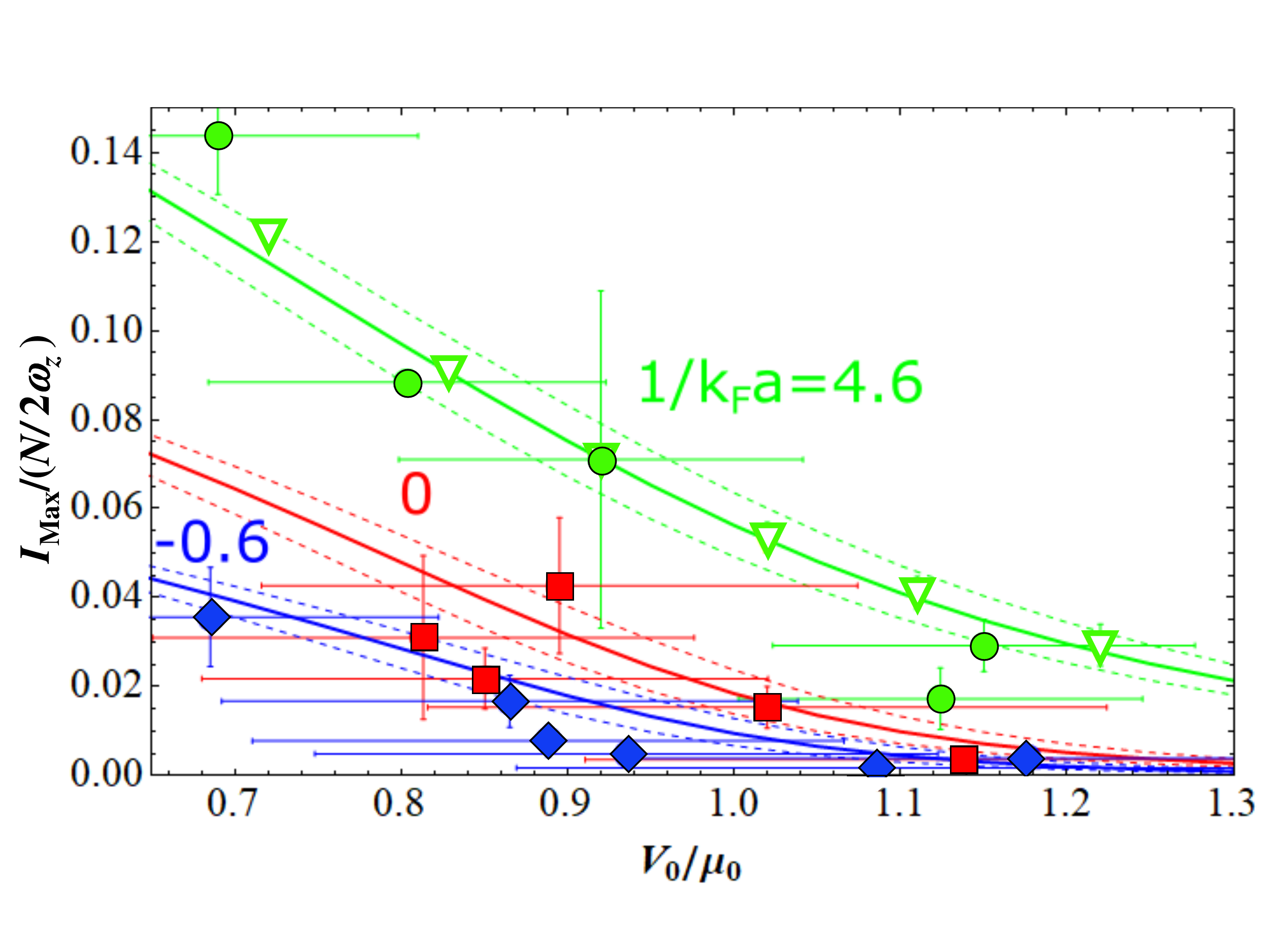}
\vskip-10pt
\caption{
Critical boson current for trapped superfluids across the BCS-BEC crossover. 
Experimentally determined $I_{Max}$ (filled symbols  plus horizontal and vertical error bars) from Ref. \cite{Burchianti2018}, 
measured as a function of $V_0/\mu_0$ for $^6$Li fermionic superfluids at unitarity ($1/k_Fa=0$, red squares), in the BEC ($1/k_Fa=4.6$, green circles) and BCS ($1/k_Fa=-0.6$, blue diamonds) limit, 
are compared with our model predictions (solid lines, see legend), employing the $\mu_B$ and $n_c$  trends obtained in the Luttinger-Ward 
approach \cite{Haussmann2007,Frank2019}.
In the BEC limit, our results are also compared with EFTM simulations \cite{Xhani2019a} (green empty triangles). 
Trap frequencies and atom number were fixed to their mean values in the experiment (nominal value in the simulation). 
The Gaussian barrier of waist $w \sim 2 \mu m$ employed in the experiment was approximated with an Eckart potential as described in the text. 
Dashed lines are our model predictions assuming a $\pm$ 5\% variation of the peak chemical potential.
}
\label{Currents}
\end{center}
\end{figure}

In Fig. \ref{Currents} we compare our theory predictions (solid lines) for the maximum currents $I_{Max}/(\frac{N}{2} \omega_z)$, plotted as a function of the barrier height $V_0/\mu_0$, with experimental data \cite{Burchianti2018} for BEC (green circles), crossover (red squares) and BCS (blue diamonds) superfluids, respectively. In the BEC limit, we also benchmark our model against results of numerical simulations \cite{Xhani2019a} (empty green triangles). 
Here $\mu_0$ denotes the pair chemical potential at the center of the cloud evaluated for the superfluid gas 
in the absence of the repulsive barrier. The different regimes are marked in the legend in terms of the peak interaction parameter $1/k_F a$.
We emphasize that, given the relatively small values of $V_0/\mu_0$ accessible in the experiments, second order effects are 
explicitly included in the evaluation of $I_{Max}$ following Ref. \cite{Goldobin2007}. 
For each interaction regime, dashed lines represent our model predictions once a $\pm 5\%$ variation of the peak chemical potential is 
assumed when evaluating the transmission amplitude.

The comparison shown in Fig. \ref{Currents} clearly demonstrates that our  model is able to reproduce the results obtained in the literature
so far with remarkable accuracy for all interactions and barrier heights. In particular, we emphasize that our analytic theory quantitatively 
matches the ETFM numerical results (green triangles) \cite{Xhani2019a} obtained in the BEC limit even for barrier heights as low as 
$V_0/\mu_0 \sim 0.7$, i.e. even when a sizable part of the trapped sample is out of the tunneling regime. There, inclusion of second order 
contributions to $I_{Max}$ is essential, as it yields corrections to the first order term up to 15\%.
Regarding the crossover and BCS regimes, our model correctly reproduces the observed trends,  at least within the large uncertainties of 
the experimental data and of the zero-temperature  approximation done in the theoretical analysis. In particular, we remark that inclusion 
of the non-perturbative results for both chemical potential and condensate fraction \cite{Frank2019} is essential to obtain quantitative 
agreement with the data, which would not match our theory if mean field results were employed.
As a consequence, although more experimental studies are required to fully validate our model within the crossover and BCS regimes, 
we stress that the measure of the critical Josephson current may provide a valuable experimental tool to determine the condensate density 
in crossover superfluids, a quantity which is hard to access by other means. 

\section{CONCLUSIONS AND OUTLOOK}

Building on a microscopic model for the Josephson dynamics between two weakly interacting Bose Einstein condensates  \cite{Meier2001}, 
we have developed a simple theoretical framework which captures the corresponding critical current of strongly interacting 
fermionic superfluids throughout the BCS-BEC crossover. 
As testified by the comparison with available numerical \cite{Zou2014,Xhani2019a} and experimental \cite{Burchianti2018}  data, 
our model provides a quantitative description of the critical currents in generic junction geometries, 
solely based on bulk properties of the superfluid state and the knowledge of the single boson transmission amplitude. 
An important feature of our theory is that it provides a consistent explanation for the non-monotonic trend for the maximum Josephson 
current observed near unitarity. Its origin is connected with the interplay between a decrease of the chemical potential and 
a concomitant increase of the condensate fraction if the interactions are tuned from the BCS to the BEC limit.
By contrast, EFTM models, to which our theory naturally connects deep in the BEC limit, fail 
when applied to intermediate and weak couplings even at the qualitative level  \cite{Spuntarelli2007, Zou2014}. 
Indeed, in these approaches the condensate density by construction coincides with the superfluid one at all couplings.  
In the future, therefore, it will be interesting to test whether a modified EFTM formalism, which explicitly accounts for 
the reduction of the condensate fraction upon going from the BEC towards the unitary gas and eventually the BCS limit, 
yields results consistent with the ones obtained here or through BdGE approaches. 
We also anticipate that our model could be extended to describe dissipative normal currents \cite{Burchianti2018}, 
and to account for finite temperature effects. 
Moreover, starting from the present study, it might be possible to investigate how the critical current evolves from the tunneling 
regime considered here to the case of weak barriers studied in Refs. \cite{Spuntarelli2007,Spuntarelli2010,Miller2007}, 
in which  pair-breaking excitations are expected to become relevant already at intermediate couplings 
close to unitarity \cite{Spuntarelli2010}.

On a rather fundamental level, it remains an open challenge to justify our use of the result in Eq. (\ref{jc2}), 
which is based on viewing the Josephson effect throughout the BCS-BEC crossover in terms of the 
coherent tunneling of pairs,  by developing a truly microscopic approach which involves the constituent fermions. 
Within such a description and in the tunneling limit, the Josephson effect is second order in the single fermion tunneling amplitudes $t_{kq}$.  Quite generally, the exact critical current at zero temperature 
\begin{equation}
I_{c} =-\frac{4}{\hbar^2} \sum_{kq}  |t_{kq}|^2 \int_0^{\infty} \frac{d\omega_1}{2\pi} \int_0^{\infty} \frac{d\omega_2}{2\pi} 
\frac{B(k,\omega_1) B(q, -\omega_2)}{\omega_1+\omega_2}
\label{IcFermi}
\end{equation}
can be expressed in terms of the spectral functions $B(k,\omega)$ associated with the anomalous propagator of fermionic superfluids.   
In the BCS-limit, $B^{\rm BCS}(k,\omega)=2\pi\, u_kv_k\, [\delta(\omega-E_k/\hbar)-\delta(\omega+E_k/\hbar)]$ 
is determined by the gap parameter $\Delta_k=u_kv_k\cdot 2E_k$ and the excitation energies $E_k=\sqrt{\xi_k^2+|\Delta_k|^2}$
of the fermionic quasiparticles in the standard manner \cite{Fetter}. The expression Eq. (\ref{IcFermi}) then reduces to the 
Ambegaokar-Baratoff result Eq. (\ref{IAB}) which involves the tunneling probability $\vert t\vert^2(\mu_F)$ evaluated right at the Fermi energy. 
In order to verify the non-monotonic dependence of the critical current on the dimensionless coupling constant $1/k_Fa$ 
along the BCS-BEC crossover within such a fermionic approach one needs to determine the anomalous spectral functions $B(k,\omega)$ that 
enter Eq. (\ref{IcFermi}) for arbitrary coupling, a problem which does not seem to have been discussed so far.
\vskip 2pt
\textit{Note Added}: After submission of this work, our model has been successfully 
employed to analyze new  experimental data by W. J. Kwon \textit{et al.}, arXiv:1908.09696, and also - in an extension to a two-dimensional 
setup - by N. Luick \textit{et al.}, arXiv:1908.09776.

\begin{acknowledgments}
We thank F. Dalfovo and P. Zou for providing the chemical potential values relevant to analyze the data of Ref. \cite{Zou2014}, 
K. Xhani for sharing the ETFM results reported in Ref. \cite{Xhani2019a} for  trapped BECs, and 
B. Frank for the non-perturbative results shown in Figure 2 and 
employed to analyze the experimental data of Figure 4.
We acknowledge fruitful discussions with F. Dalfovo, E. Neri, A. Recati, G. Roati, F. Scazza and K. Xhani. 
This work was supported by the ERC through grant no.\:637738 PoLiChroM 
and in part also by the Nano-Initiative Munich (NIM). 
\end{acknowledgments}

\vskip50pt


\begin{thebibliography}{53}
\bibitem [{foo()}]{footnote}%
  \BibitemOpen
  \href@noop {} {}\bibinfo {note} {Formally, $\varphi=\varphi_1-\varphi_2$ is
  determined by the phases associated with the anomalous expectation values
  $\langle\hat{\Psi}\rangle_{1,2} =\sqrt{n_c}e^{i\varphi_{1,2}}$ of the
  condensates on both sides. It is nonvanishing only in the presence of an
  externally imposed current $I(\varphi)\ne 0$.}\BibitemShut {Stop}%
\end{thebibliography}

%

\end{document}